\documentclass[pre,eps,aps,reprint,twocolumn,superscriptaddress]{revtex4-2}
\usepackage{graphicx}
\usepackage{amsmath}
\usepackage{amssymb}
\usepackage{color}
\usepackage[latin2]{inputenc}
\def\slaninafigdir{.}
\begin{document}
\title{%
Ratchet effect and jamming in dense mixtures of active and passive colloids in narrow pores
}
\author{%
Franti\v{s}ek Slanina%
}%
\affiliation{%
Institute of Physics,
Czech  Academy of Sciences, 
 Na~Slovance~2, CZ-18200~Praha,
Czech Republic%
}%
\email{
slanina@fzu.cz
}%
\author{%
Miroslav Kotrla%
}%
\affiliation{%
Institute of Physics,
Czech  Academy of Sciences,
 Na~Slovance~2, CZ-18200~Praha,
Czech Republic%
}%
\begin{abstract}
Using the framework of generalized exclusion processes we study
mixtures of passive and active particles interacting by steric
repulsion. The particles move in a pore with periodically modulated
aperture, which is modeled by a quasi-one-dimensional channel with
periodic tooth-shaped profile. Internal driving of the active
particles induces a ratchet current of these particles. In the
current-density diagram, we observe three main regimes: of free flow;
of thermally activated processes; of spinodal decomposition.
When the
density of particles is increased, we observe a transition to jammed
state, where the ratchet current is substantially reduced.
In time evolution, the transition to jammed state is seen as
sudden drop of current at certain time. The probability distribution
of these jamming times follows an exponential law. The
average jamming time depends
itself exponentially on the density of active particles. The
coefficient in this exponential is nearly independent of the switching
rate of the active particles as well as on the presence or absence of
passive particles.
Due to
the interaction, the current of active particles imposes a drag on the
passive particles. In the limit of both large systems and long times,
the current of passive particles has always the same direction as the
ratchet current of active particles. However, during the evolution of
the system we observe very slow
(logarithmic in time) approach to the asymptotic
value, sometimes accompanied by current reversal, i.e. current of
active and passive particles may go in opposite direction.
\end{abstract}
%
%
\maketitle%
\section{Introduction}

In microfluidic applications
\cite{squ_qua_05,whitesides_06,huber_15} one often encounters
multiphase flow, be it microscopic bubbles, droplets or solid
colloidal particles carried by the host fluid. Here we focus on
suspensions of colloid particles flowing through constrained
environments
\cite{ket_rei_han_mul_00,mar_bug_tal_sil_02,mat_mul_03,han_mar_08,dic_edd_hum_sto_ton_09,bur_han_mar_sch_tal_09,cis_vas_par_and_11,yan_liu_li_mar_han_zha_17,how_gau_pan_nik_16,sol_ara_her_17},
namely through narrow pores. One of the important effects which appear
in such geometry is the emergence of ratchet current \cite{reimann_02}.  It originates
from unbiased periodic driving on condition that the forward-backward
symmetry of the pore geometry is broken. In the classical experiment
\cite{mat_mul_03,mat_mul_gos_11} micrometer-sized spherical particles acquired
rectified flow through silicon wafer pierced by etched pores of wavy
profile. This is the geometry that we have in mind when setting up our
model.

As the ratchet current depends sensitively on size and other
properties of the colloidal particles, the ratchet effect is a natural
tool in sorting the mixtures of particles of various types
\cite{saj_sen_14,xua_lee_14,sal_zem_zha_17,gho_han_mar_mar_nor_sch_sch_12,ver_gri_kre_tan_kan_erb_sch_maa_12,sch_fri_due_ryu_kno_18}.
In our previous works, we investigated the sorting capacity of such
ratchet devices both in dilute (i.e. effectively one-particle) regime
\cite{slanina_20,sla_kal_19,slanina_19,slanina_16} and in dense
(i.e. nearly full-packed) regime
\cite{hum_kot_net_sla_20,sla_kot_net_22,sla_kot_23}. In this work we complement
our results by considering one component of the colloidal mixture
composed of active particles
\cite{rom_bar_ebe_lin_12,fod_mar_18,ramaswamy_10,mar_joa_ram_liv_pro_rao_sim_13,pro_jul_joa_15,bec_dil_low_rei_vol_vol_16}.
Indeed, while the ratchet effect in
passive Brownian particles is driven by alternating external filed,
the same role is assumed by internal driving which is individual for
each active particle separately.

Ratchet effect in ensembles of active particles was already intensely
studied \cite{ols_rei_17}, experimentally
\cite{gal_key_cha_aus_07,hul_dil_she_etal_08,mah_cam_bis_kom_cha_soh_hud_09,dil_ang_del_ruo_10,sok_apo_grz_ara_10,vol_but_vog_kum_bec_11,bri_cau_des_dau_bar_13}
as well as theoretically,
both in 2D \cite{wan_ols_nus_rei_08,ang_dil_ruo_09,rei_rei_13,ber_etal_13,gui_jey_berd_etal_14}  and 1D
\cite{ang_cos_dil_11,ai_che_he_li_zho_13,pot_hah_sta_13,gho_mis_mar_nor_13,gho_han_mar_nor_14,ao_gho_li_sch_han_mar_14,kou_mag_dil_14,ai_he_zho_14,ao_gho_li_sch_han_mar_15,ai_16,ai_he_zho_17,rubi_19,bis_mar_20}
geometries.

Macroscopic ensembles of active particles are inherently
far-from-equilibrium systems. Therefore, effects forbidden by
equilibrium thermodynamics can occur. One of them is motility induced
phase separation (MIPS), or dynamical
freezing \cite{rei_rei_11,bia_low_spe_13,cat_tai_15,dig_lev_sum_cug_gon_pag_18,cap_dig_lev_cug_gon_20},
which can be observed even in
one-dimensional systems  \cite{tai_cat_08}. One of the main questions
we want to address in this work is how the ratchet effect interferes
with dynamical freezing. The ratchet effect, as such just a
one-particle phenomenon in quasi-1D confinement, continues to be at
work when many particles
start to interact with each other. Intuitively, the interactions
hinder the movement of particles and thus suppress the ratchet
effect. This can be visualized in the current-density diagram, as shown
e.g. in \cite{slanina_09} for the case of softly interacting
Brownian motors in one dimension. For passive but driven particles this behavior is
generic and finds its paradigmatic example in the asymmetric simple
exclusion process (ASEP). Therefore, it is natural to start
investigating the effect of active, rather than passive, interacting
particles in quasi-1D geometry by generalizations of
ASEP. This way was already undertaken several times
\cite{sot_gol_14,slo_eva_bly_16,mal_bly_eva_19,rav_ang_20,sch_row_mal_23}. Two-dimensional
models of this type were  studied in the context of active lattice gases
\cite{pil_eav_14,sol_tai_15,man_pug_17,kou_eri_bod_tai_18,dit_spe_vir_21,mas_eri_jac_bru_23}.

In this work, we try to apply similar approach to asses the influence the
dynamical freezing has on the ratchet current. Specifically, how the
ratchet effect, seen as a vehicle for particle sorting, works in
circumstances affected by the dynamical freezing. To this end, we
introduce a model based on generalized ASEP \cite{sla_kot_23} in which
a mixture of active and passive particles\cite{dec_dil_sot_sol_21}
moves in quasi-1D pore of
a tooth-shaped profile. We expect that the ratchet effect, which in
such mixture affects directly only the active component, is imprinted
via the interaction also on the passive particles. How big such effect
is, how it is affected by dynamical freezing and whether there are any
sign changes in the current of passive particles, that is to be
answered by our simulations.

\section{Active and passive particles in tooth-like pore}

We assume active and passive particles performing Brownian motion. The
active component is also subject to active drift and angular
diffusion. All particles
interact by steric repulsion. To account for the interaction, we apply
a scheme called by us ``local mixing approximation'' in our previous
works \cite{sla_kot_net_22,sla_kot_23}. It consists in discretizing
the space into cells and imposing a constraint on number of particles
which can simultaneously occupy one cell. Suppose we have altogether
$M$ species of particles. Particles of species $\sigma$ are
characterized by number $d_\sigma$ (we call it size) and there are
$n_\sigma$ of them in a chosen cell. The cell as a whole is
characterized by its capacity
$k$. We implement the steric repulsion of the particles by
requiring that the inequality $\sum_{\sigma=1}^Md_\sigma n_\sigma \le
k$ holds for all cells. In this work, we shall use just two species of
particles. The particles we shall call ``small'' have size $d_s=1$,
the particles we shall call ``big'' are twice as large, $d_b=2$. The
cell capacity in all what follows is fixed as $k=3$. Therefore, we
impose the constraint
\begin{equation}
n_s+2n_b\le 3
\label{eq:constraint}
\end{equation}
in all cells.

\begin{figure}[t]
\includegraphics[scale=0.3]{%
\slaninafigdir/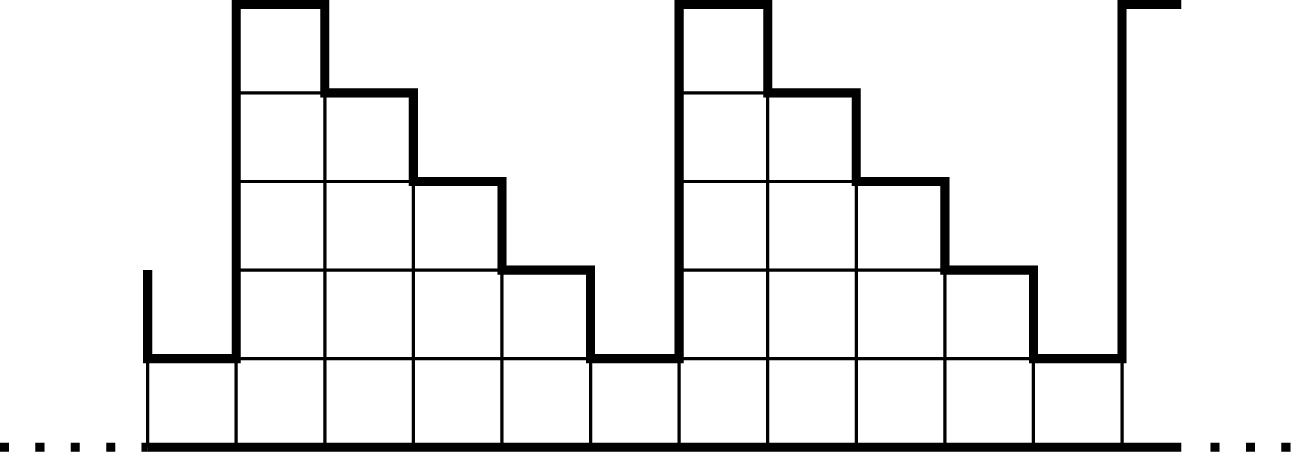}
\caption{Scheme of the geometry of the pore, consisting of
periodically repeated teeth. Each tooth is composed of
$15$ square cells drawn by the thin lines. Number of small ($n_s$)
and big ($n_b$)
particles in each of these cells must satisfy the condition
$n_s+2n_b\le 3$.
}
\label{scheme-zuby}
\end{figure}

In order to exhibit the ratchet effect, the quasi-1D pore must break
the backward-forward mirror symmetry. As a typical example of such
geometry we choose periodic array of two-dimensional teeth, as shown
in Fig. \ref{scheme-zuby}. Each tooth is composed of $15$ square
cells, and the condition  $n_s+2n_b\le 3$ must hold in each of these
cells. In simulations, we suppose that the system has $L$ teeth
(i. e. $15L$ cells) and periodic boundary conditions apply. In the
system, there are $N_s$ small particles and $N_b$ big
particles. Average density of small and big particles is
$\rho_s=N_s/(15L)$ and $\rho_b=N_b/(15L)$, respectively. The local
condition  (\ref{eq:constraint})  implies the global one
$\rho_s+2\rho_b\le 3$.

\begin{figure}[t]
\includegraphics[scale=0.45]{%
\slaninafigdir/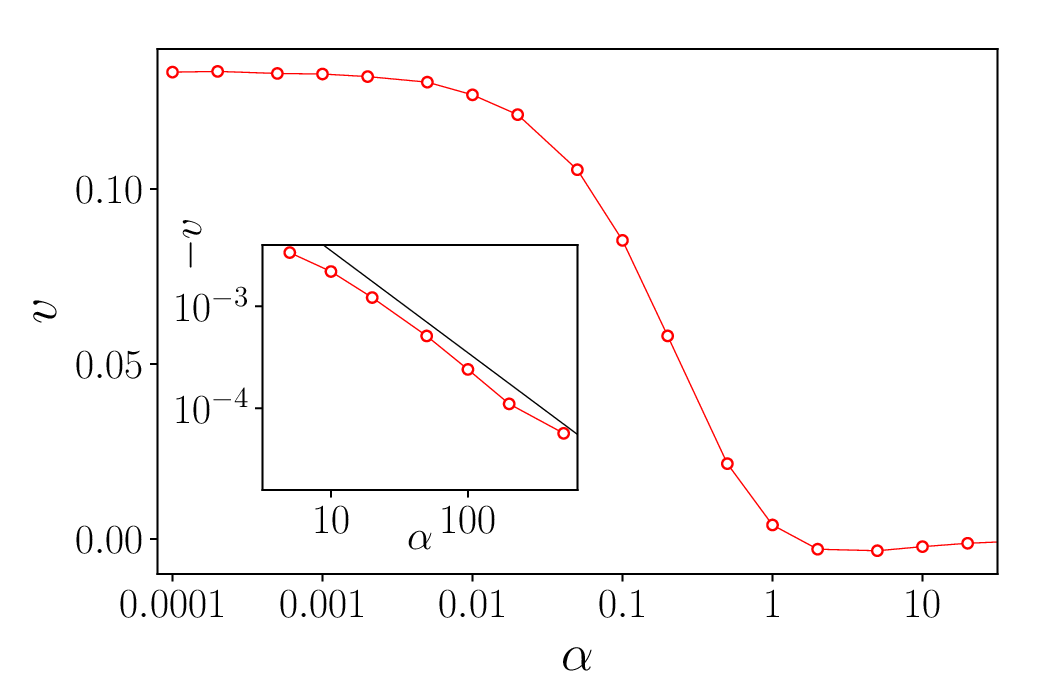}
\caption{Dependence of the ratchet velocity of small particles on the
switching rate, in the
limit of small density $\rho_s\to 0$. In the inset we show the
approach of the velocity to zero when $\alpha\to\infty$. The
straight line is the dependence $\propto 1/\alpha$. Note that the
velocity changes sign around $\alpha\simeq 1.5$ and becomes negative
for large $\alpha$.
}
\label{velocity-vs-alpha-L1000-Nb0}
\end{figure}

The dynamics of the system is a sequence of jumps of individual
particles. Each particle can jump to any of its four (or less, if on
the boundary) nearest neighbors, on condition that
(\ref{eq:constraint}) is satisfied everywhere at all times. Let us now
describe the jump rates. They depend on the species of the particle
and on the direction of the attempted jump.

We suppose that the small particles are active and the big ones
are passive. So, the jump rates of big particles are isotropic and the
particle attempts to jump to each of its four neighbors with equal
rate $A$. The small particles have each its internal direction, which
can assume four values  $\zeta\in\{0,1,2,3\}$, where, conventionally,
we denote the directions right, up, left, down, by numbers $0,1,2,3$,
respectively. The same convention holds for the direction $\xi$ of the
attempted jump. Now, the rate of jumps in direction $\xi$ for particle
with internal direction $\zeta$ is $r(\zeta+\xi\mod 4)$ and we denote
the four rates
$[r(0),r(1),r(2),r(3)]=[a,c,b,d]$ (compatible with our earlier work
\cite{sla_kot_23}).  Moreover, the internal direction of the small
particles also jumps among its four values as
$\zeta\to\zeta'=\zeta\pm 1 \mod 4$ with rate $\alpha$ (we shall call
it the switching rate).

By symmetry, we must have $c=d$. Because the Brownian motion is
assumed isotropic even for active particles, we have
$a+b=c+d$. Moreover, the latter number just fixes the unit of time,
so, there is only one independent physical parameter characterizing
the small particles, which is $a-b$, the active drift. As for the big
particles, we assume $A=a/2$, according to the generic result that
diffusion constant of Brownian particles is inversely proportional to
their size. (Precise relation of $a$ and $A$ would require much deeper
study, though, because the relation of ``sizes'' $d_b=2d_s$ is just a
rough approximate relation between the diameters of the big
and small particles. We do not want to dig into this field here.)

We study the model by direct numerical simulations.
 {In the practical implementation, in each step the
algorithm picks a particle at random and then chooses among six
possible events, namely change of internal direction, or attempt of
hopping in four possible directions, or no change. Probabilities of
the first five events are chosen so that they are
proportional to prescribed rates $\alpha$ and $a$, $b$, $c$,
$d$ and at the same time their sum is at most one, in order to have
the sixth probability (of no change) non-negative. Otherwise, we
have freedom in choosing these probabilities. However, a care must
be taken, because the units of time must be rescaled accordingly, in order to
get correct values of currents.}

We mostly
fix the length of the pore $L=1000$, except when finite-size effects
are investigated. We also fixed the jump rates $a=1.5$, $b=0.5$,
$c=d=1$, $A=0.5$ throughout, as the only one physically relevant free
parameter among the set of hopping rates is the active drift $a-b$ and the influence of its precise value is
marginal compared to the strong dependence on the switching rate $\alpha$ and
especially on the densities $\rho_s$ and $\rho_b$.

\begin{figure*}[t]
\includegraphics[scale=0.7]{%
\slaninafigdir/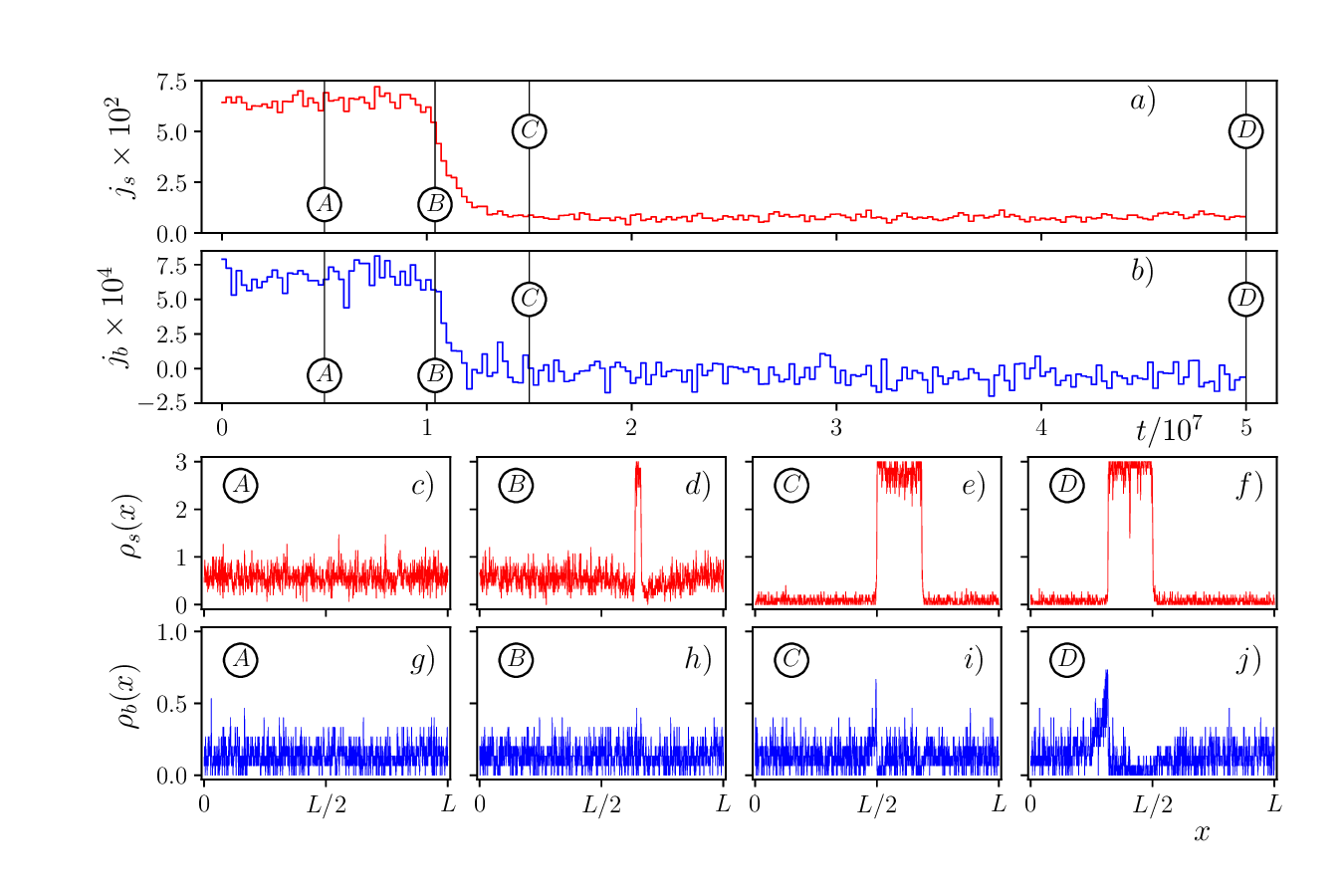}
\caption{Time evolution of the current of small (panel a)) and big
(panel b)) particles, and snapshots of local density of small
(panels c) to f)) and big (panels g) to j)) particles. The moments
at which the snapshots are taken are indicated in panels a) and b)
by letters $A$, $B$, $C$, and $D$. The length of the system is
$L=1000$, the number of small particles is $N_S=8700$, the number of
big particles is $N_b=2000$ and the switching rate
$\alpha=0.001$. The coordinate $x=1,2,\ldots,L$ numbers the teeth
along the pore.  The values $\rho_s(x)$ and $\rho_b(x)$ are the
average densities of small and big particles, respectively, in the
tooth number $x$.
}
\label{configurations-L1000-Ns8700-Nb2000-alpha0p001}
\end{figure*}

\section{Free flow, metastable and jammed states}

As there is no explicit global drive, the net average current of small
particles $j_s$ as well as big particles $j_b$ is entirely due to the
ratchet effect. Let us first focus on the simpler situation where
density of small (active) particles is so small, that their mutual
interaction can be neglected.   We have effectively a one-particle
problem and we can measure the average ratchet velocity
of the small
particles
\begin{equation}
v=\lim_{\rho_s\to 0}j_s/\rho_s\;.
\end{equation}
In such case, the presence of big
(passive) particles is irrelevant, because the cell capacity is
assumed $k=3$ and there is always free space for one small
particle in any cell, irrespectively of whether there is also a big particle in
the cell or not. We show in Fig. \ref{velocity-vs-alpha-L1000-Nb0} the
dependence of the ratchet velocity on the switching rate $\alpha$. As
we can see, for $\alpha\to 0$ it approaches a limit which corresponds
to adiabatic approximation. Here the adiabatic approximation consists
in assumption that the system is a mixture of four species of small
particles, each moving in one of the four allowed directions
forever. For increasing $\alpha$ the ratchet effect is suppressed and
the ratchet velocity eventually tends to zero. In our geometry, the current also
changes sign around $\alpha\simeq 1.5$ and approaches zero as
$-v\propto 1/\alpha$ for $\alpha\to\infty$. Such behavior is
consistent with earlier studies of single-particle ratchet effect,
which are based on $1/\alpha$-expansions
\cite{kal_sla_21,kalinay_22,kal_sla_23}.

When the density of small particles is increased, the interactions
start to play a role. Up to certain value of the density, the
interactions just
slightly reduce the current. However, beyond certain limit, the
behavior changes dramatically, due to appearance of jams that hinder
the current substantially. These jams appear as a consequence of
dynamical freezing, common in the ensembles of active particles
\cite{tai_cat_08,rei_rei_11,cat_tai_15}. We illustrate in
Fig. \ref{configurations-L1000-Ns8700-Nb2000-alpha0p001} the typical
behavior. We show there time dependence of the current of both small
and big particles and
snapshots of local density of particles (small as well as big) at four
moments which represent well the typical configuration of particles at
different phases of the time evolution. Here and in the following, the
time is measured in units of simulation sweeps, each consisting of $L$
elementary moves, if $L$ is the system length (the number of
teeth). Initially, the current
stabilizes and fluctuates around a quasi-stationary state. The density
is uniform (up to fluctuations), as shown in snapshots $A$. By using
the prefix ``quasi'' we intended the fact that after some time (in
Fig. \ref{configurations-L1000-Ns8700-Nb2000-alpha0p001}  it is
around time $t=10^7$) the jam starts forming and the currents of
both small and big particles drop simultaneously to significantly
smaller value. In the snapshots $B$ we can see how the jam starts
forming. Initially, a small nucleus appears, in which the density of small
particles approaches its maximum value
$\rho_{s}^{\mathrm{max}}=3$. In the vicinity of the nucleus, the density
of small particles is
suppressed, but in the rest of the system the density remains as in
the snapshot $A$. During the formation of the jam, the current steadily
decreases, but ultimately it stabilizes at (now truly) stationary
value. Fully developed jam can be seen in snapshots $C$. It can be
roughly described as follows. The jam consists of a compact area
in which the average density of small particles is close to its maximum value
$\rho_{s}^{\mathrm{jam}}\simeq\rho_{s}^{\mathrm{max}}$, while in the rest of the system the density is
uniformly reduced to an ``unjammed'' value $\rho_{s}^{\mathrm{unjam}}$.
The presence of the jam is not much visible in the
density of big particles, but still, a peak of the density of big
particles can be seen just at the edge of the jam. This can be
understood as blocking the big particles from entering the area of the
jam, but the big particles which were in this area when the jam was
formed, remain there. Finally, we may look how the jam evolved after
some time after its formation. This is shown in snapshots $D$. We can
see that the structure of the
jam remains essentially unchanged, only its position is shifted. The
direction of this shift is opposite with respect to the direction of
the ratchet current of small particles. Indeed, once the jam is
formed, it evolves by absorbing particles coming from the left
(assuming, as in the figure, ratchet current oriented left-to-right)
and ``evaporating'' particles at its right edge. This results in
leftward shift of the jam. We also observe small change in the density
of big particles. The peak in density of big particles at the left
edge of the jam is now higher and the density of big particles within
the jam is smaller than in the snapshot $C$. It means that during the
leftward movement of the jam, the big particles are slowly wiped out
of the jam interior. Simultaneously, they are accumulated at the edge
of the jam.

The fundamental observation is that at given values of the average
densities  $\rho_s$ and $\rho_b$ there are two characteristic values
of the current of small particles $j_s$ (and analogously, two  characteristic values
of the current of big particles $j_b$). Indeed, we can measure the quasi-stationary
current before the jam is formed (in the following we shall call this
state a metastable state and the current a metastable current). We can
also wait until the jam is formed and only after that start measuring
the true stationary current which persists in the jammed state. Both
of these values can be measured in
the same simulation run, but it may also happen that either the jam is
never observed or the jam occurs so early in the simulation that the
metastable state cannot be observed. Typically, the first case is
found for low enough $\rho_s$, while the second for large enough
$\rho_s$. Both metastable and stationary currents are observable only
in a limited interval of densities in the middle, as we shall see
later.

From the simulations it is difficult to establish with certainty, if
the absence of jams at a certain low value of the density is due to too short
simulation run or it is a sign of specific phase without jamming at
arbitrarily long times. Here we hypothesize that such unjammed phase
does really exist and we shall call it the free-flow phase.  On the
other hand we found it rather difficult to establish the value
of the density at which the jamming first occurs, i.e. the point of
jamming transition. Later we shall provide some hints how to estimate
the jamming point indirectly.

The presence of well-distinguishable metastable and stationary states
implies that the current-density diagram will consist of two separate
branches. In certain interval of the densities the two branches will
lie one above the other and there will be two values for current for
each density. This feature is reminiscent of a hysteresis curve and
the jamming-unjamming transition exhibits similarity with first-order
equilibrium phase transitions.

\begin{figure}[t]
\includegraphics[scale=0.45]{%
\slaninafigdir/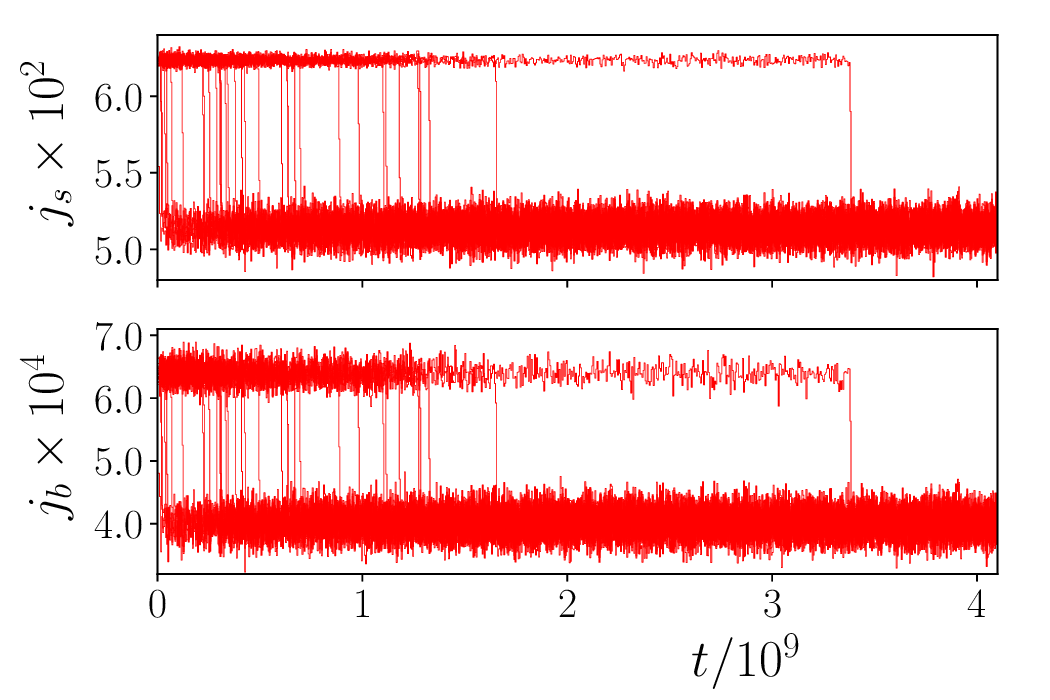}
\caption{Time evolution of the current of small (upper panel) and big
(lower panel) particles. Each line represents a single
independent run. There are $36$ runs shown in this figure. The length of
the system is $L=1000$ and contains
$N_S=8700$ small and $N_b=2000$ big particles. The switching rate is
$\alpha=0.01$.
}
\label{evolution-jamming-L1000-Ns8700-Nb2000-alpha0p01}
\end{figure}

\begin{figure}[t]
\includegraphics[scale=0.45]{%
\slaninafigdir/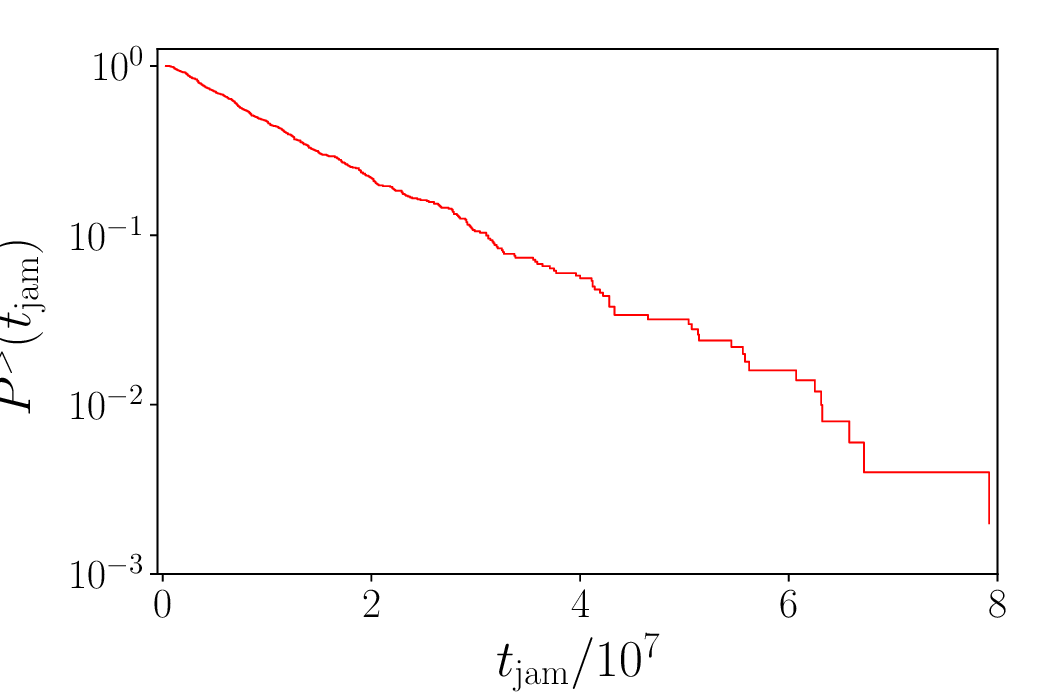}
\caption{Cumulative distribution of jamming times for system of length
$L=1000$, containing $N_s=11500$ small active particles and
$N_b=500$ big passive particles. The switching rate is $\alpha=0.01$.
}
\label{distr-jamming-time-L1000-Ns11500-Nb500-alpha0p01}
\end{figure}

Besides the values of the metastable and stationary currents, we are
interested also in times at which the jam starts to form, i.e. the jamming
times $t_\mathrm{jam}$. It turns out that
jamming times fluctuate wildly from one realization of the simulation to
another. We show in
Fig. \ref{evolution-jamming-L1000-Ns8700-Nb2000-alpha0p01} time
evolution of the currents of small and big particles for many
independent realizations. For all densities  we
investigated and for switching rates $\alpha\le 0.01$,
the probability distribution of jamming times follows an
exponential law, an example of which is shown in
Fig. \ref{distr-jamming-time-L1000-Ns11500-Nb500-alpha0p01}.
For switching rates larger than about $0.02$ we found it
difficult to establish precisely the jamming time, because the
metastable and stationary currents are too close to each other, within
the range of stochastic fluctuations of the current. Moreover, we
found that for $\alpha \gtrsim 0.02$ the jammed state not always lasts
for the whole duration of the simulation after the jam is formed, but
occasionally the jam dissolves and the current returns to its
metastable value. Therefore, the jam is characterized by a finite
lifetime. Here we shall concentrate on the regime of lower $\alpha$,
for which the lifetime of the jam can be considered infinite, at least
compared to the maximum simulation time used. In such regime, the
set of jammed configurations are viewed as a trap of
effectively infinite depth and
further evolution is possible only within such set of
configurations.

\begin{figure}[t]
\includegraphics[scale=0.45]{%
\slaninafigdir/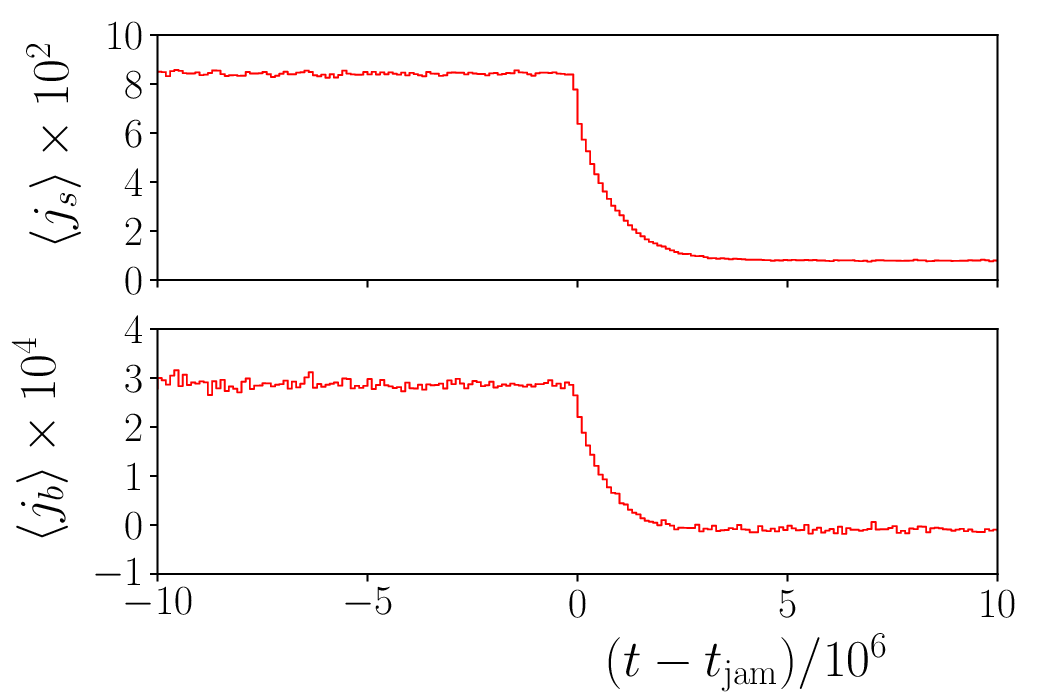}
\caption{Averaged current of small active (upper panel) and big
passive (lower panel) particles before and after the onset of the
jamming. The averaging is over $150$ independent runs. The actual
time dependence of the current in each run is shifted by the jamming
time $t_\mathrm{jam}$ which in this figure is fixed as the time at
which the current of small particles drops below the value $j_s=0.07$.
The  length of the system is $L=1000$ and contains
$N_s=10700$ small (i.e. $\rho_s=0.7133\ldots$) and $N_b=500$ big
particles. The switching rate is
$\alpha=0.001$. }
\label{evolution-jamming-posunute-L1000-Ns10700-Nb500-alpha0p001}
\end{figure}

We
have seen already in
Fig. \ref{configurations-L1000-Ns8700-Nb2000-alpha0p001} that the
formation of
the jam takes some
time. In order to
see better the evolution during the formation of the jam,
we shifted the time evolution of the current  so that the
drop initiates  at time zero, and then averaged it over many
independent runs. The result is shown in
Fig. \ref{evolution-jamming-posunute-L1000-Ns10700-Nb500-alpha0p001}
and suggests the following scenario. We
can see a steep initial decrease of the current at the moment when the
jam first appears. Then, the current approaches more slowly to the
stationary value as the jam gradually grows and the density of particles outside
the jam decreases to the value $\rho_{s}^{\mathrm{unjam}}$. Note that the
time $t_\mathrm{form}$ for the formation of the jam can be relatively
long, in this case $t_\mathrm{form}\simeq 10^6$, which is of the same
order of magnitude as the average jamming time, in this case
($\rho_s=0.7133\ldots$, $\alpha=0.001$, $N_b=500$) we have
$\langle t_\mathrm{jam}\rangle=5.9\cdot 10^6$. However, the initial
decrease in current is indeed sharp, which enables to determine the
jamming time in each individual run with sufficient precision.

\begin{figure}[t]
\includegraphics[scale=0.45]{%
\slaninafigdir/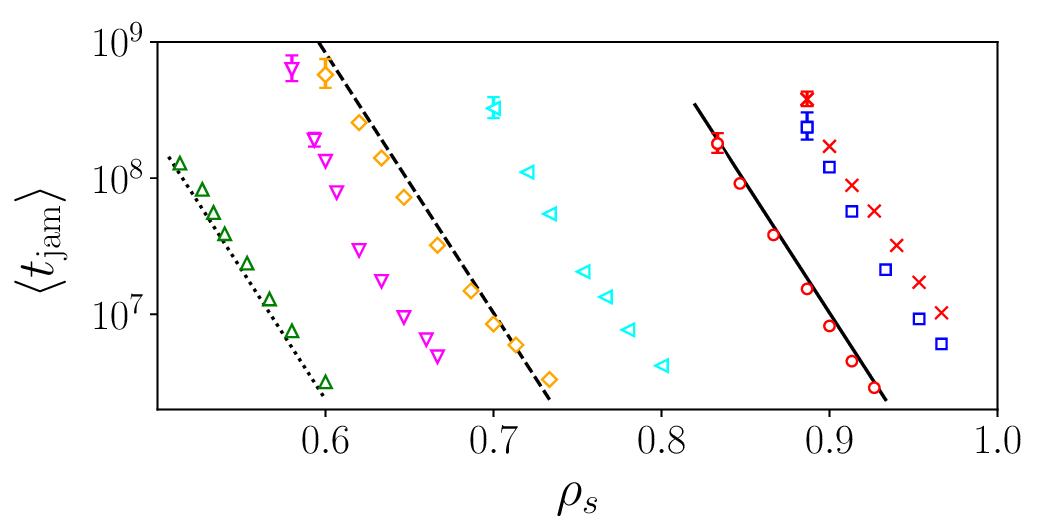}
\caption{ {Dependence
of average jamming time  on the average density of small particles. The
length of the system in $L=100$ (symbol $\times$) and
$L=1000$ (all other symbols). The number of big particles is $N_b=0$
(symbols {\Large $\circ$}, {\Large $\bullet$}, $\square$,
$\blacksquare$), $N_b=500$ (symbols  {\Large$\diamond$}, {$\blacklozenge$},
{\Large$\triangleleft$},   {\large$\blacktriangleleft$}), $N_b=2000$
(symbols $\bigtriangleup$, {\large$\blacktriangle$},
$\bigtriangledown$,  {\large$\blacktriangledown$}).
The switching rate is $\alpha=0.001$ (symbols  {\Large $\circ$},
{\Large $\bullet$},  {\Large$\diamond$}, {$\blacklozenge$},
$\bigtriangleup$, {\large$\blacktriangle$}, $\times$,$+$) and
$\alpha=0.01$ (symbols  $\square$,   $\blacksquare$,
{\Large$\triangleleft$},
{\large$\blacktriangleleft$},   $\bigtriangledown$,
{\large$\blacktriangledown$}).
The three straight lines are the exponential functions
$K   e^{-\kappa\rho_s}$ where $\kappa=44.0$ is the same in all
three cases, while $\ln K=55.7$ (full line),  $46.9$ (dashed
line), and $41.1$ (dotted line).
Where the error bars are not shown, they are
smaller than the symbol size.}
}
\label{stopping-time-vs-rhos}
\end{figure}

The most important question is how the jamming times and both
metastable and stationary current depend on the density of particles.
The exponential distribution of jamming times
suggests that the set of jammed configurations functions as a trap in
the whole configuration space and the time constant of the exponential
distribution is just the waiting time for the system to fall into such
trap. This waiting time is just the average
jamming time $\langle t_\mathrm{jam}\rangle$. We show its dependence
on the density of small particles for several values of the number of
big particles and switching rates in
Fig.
\ref{stopping-time-vs-rhos}.
Surprisingly enough, the dependence
is exponential
\begin{equation}
\langle t_\mathrm{jam}\rangle \simeq K e^{-\kappa \rho_s}
\label{eq:stoppingtime}
\end{equation}
with the value of the coefficient
 {  $\kappa\simeq 44.0$
almost independent of
$\alpha$ and $N_b$. We found that it is also independent of the system size.
The factor $K$ decreases with increasing number of
big particles  and increases with increasing switching rate $\alpha$.}
The slight deviation from the exponential law (\ref{eq:stoppingtime})
which can be seen in Fig.
\ref{stopping-time-vs-rhos}
occurs at
such densities where the values of currents in metastable state and
stationary state come close to each other. At the hypothetical
crossing point a true transition from unjammed to jammed phase is
expected, accompanied by divergence of time scales, in our case by
expected divergence of the average jamming time $\langle
t_\mathrm{jam}\rangle$. However, far from this hypothetical transition
point the dependence  (\ref{eq:stoppingtime}) seems to hold
universally.

\begin{figure}[t]
\includegraphics[scale=0.45]{%
\slaninafigdir/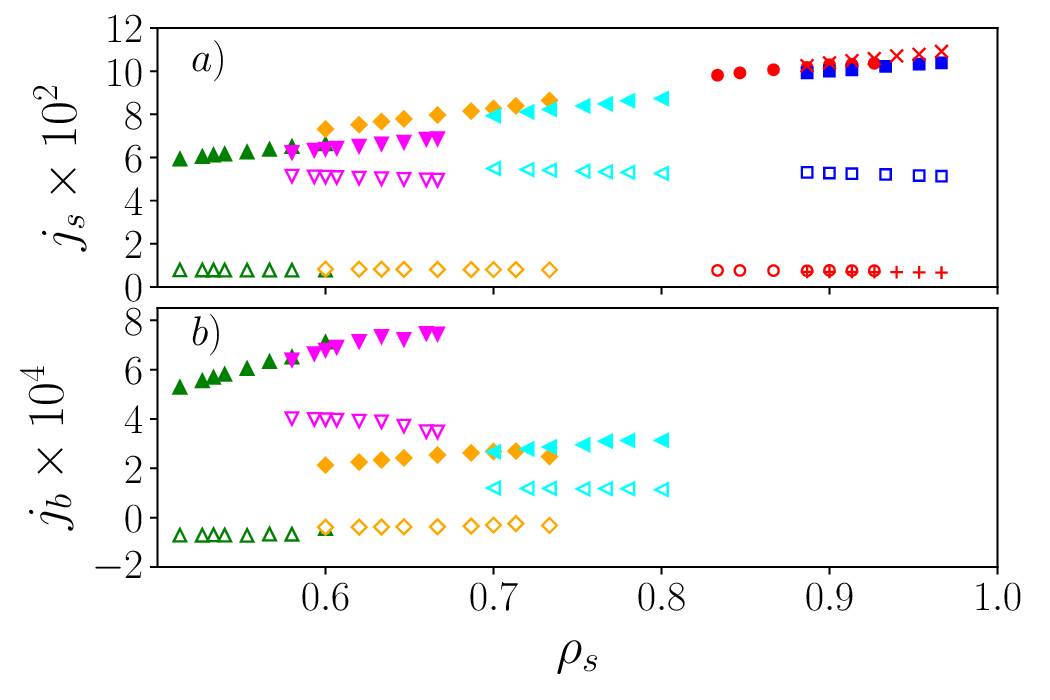}
\caption{ {Dependence of the current of
small particles (panel a)) and big
particles (panel b)) on the average density of small particles. The
length of the system in $L=100$ (symbols $\times$ and $+$) and
$L=1000$ (all other symbols). The number of big particles is $N_b=0$
(symbols {\Large $\circ$}, {\Large $\bullet$}, $\square$,
$\blacksquare$), $N_b=500$ (symbols  {\Large$\diamond$}, {$\blacklozenge$},
{\Large$\triangleleft$},   {\large$\blacktriangleleft$}), $N_b=2000$
(symbols $\bigtriangleup$, {\large$\blacktriangle$},
$\bigtriangledown$,  {\large$\blacktriangledown$}).
The switching rate is $\alpha=0.001$ (symbols  {\Large $\circ$},
{\Large $\bullet$},  {\Large$\diamond$}, {$\blacklozenge$},
$\bigtriangleup$, {\large$\blacktriangle$}, $\times$,$+$) and
$\alpha=0.01$ (symbols  $\square$,   $\blacksquare$,
{\Large$\triangleleft$},
{\large$\blacktriangleleft$},   $\bigtriangledown$,
{\large$\blacktriangledown$}).
The symbol $+$
and all empty symbols ( {\Large $\circ$},  $\square$,
{\Large$\diamond$},  {\Large$\triangleleft$}, $\bigtriangleup$,
$\bigtriangledown$)  denote the value of stationary current after the
jamming occurred; the symbol $\times$ and all filled symbols
({\Large $\bullet$}, $\blacksquare$, $\blacklozenge$,
{\large$\blacktriangleleft$},
{\large$\blacktriangle$}, {\large$\blacktriangledown$}) denote the
value of the metastable current before the jamming sets on.}
}
\label{js-jb-vs-rhos}
\end{figure}

We can see in Fig.
\ref{js-jb-vs-rhos}a
both the metastable and stationary current of small particles for the
same densities and switching rates as used for the average jamming
times.  We can see that the dependence on the number of
big particles is very weak both in metastable and in stationary
current. Also the dependence on the switching rate is barely visible
in the metastable current. On the contrary, the influence of the
switching rate on the stationary current is dramatic.
Similarly, we show in
Fig.
\ref{js-jb-vs-rhos}b
the metastable and stationary
current of big particles. Again, we can see that the metastable current is nearly
independent of the switching rate. The stationary current shows more
complicated behavior, which we shall describe later.


The data shown in Figs. \ref{stopping-time-vs-rhos} and
\ref{js-jb-vs-rhos} cover only
such intervals of the density  $\rho_s$ in which both metastable and
stationary currents can be measured and therefore the dependence of
current on density is a two-valued function. For lower densities, we
observe just the continuation of the upper branch, for higher densities just
the continuation of the lower branch. The full current-density
diagram has therefore two disjoint branches.

\begin{figure}[t]
\includegraphics[scale=0.45]{%
\slaninafigdir/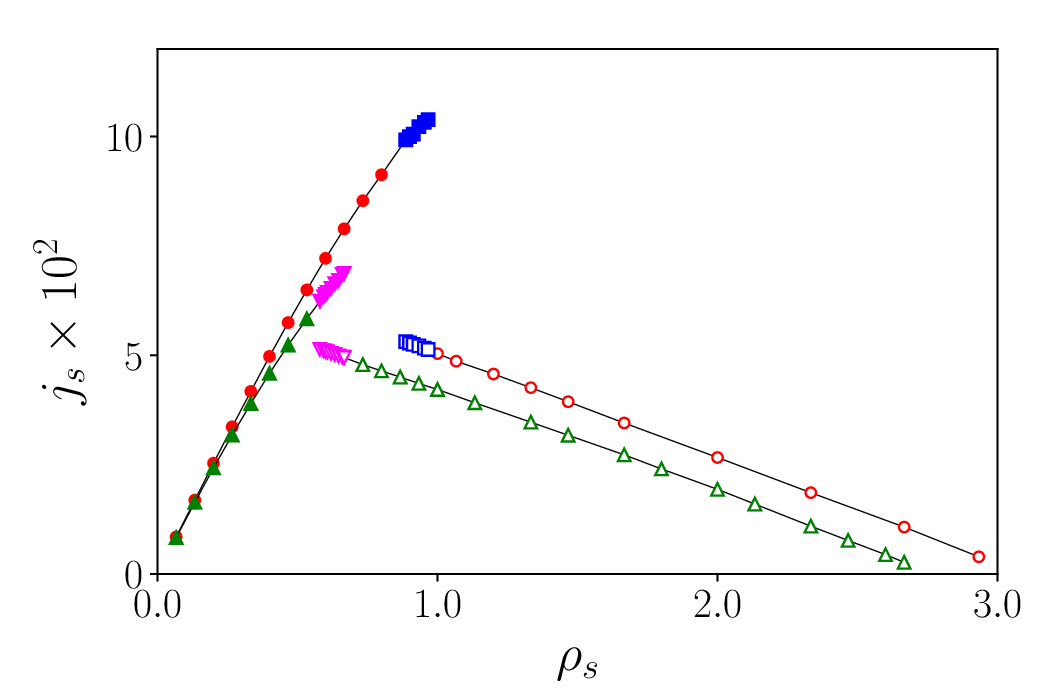}
\caption{Dependence of the current of small particles on the density
of small particles. The length of the system is $L=1000$, the
switching rate is $\alpha=0.01$, the number of big particles is
$N_b=0$ (symbols {\Large$\circ$}, {\Large $\bullet$}, $\square$, $\blacksquare$)
and $N_b=2000$ (symbols $\bigtriangleup$, {\large$\blacktriangle$}
$\bigtriangledown$, {\large$\blacktriangledown$}).  The symbols
$\blacksquare$ and {\large$\blacktriangledown$} denote metastable
current before jamming occurs, the symbols  $\square$ and  $\bigtriangledown$
denote
stationary current after jamming.
}
\label{js-vs-rhos-L1000-Nb0Nb2000-alpha0p01}
\end{figure}

\begin{figure}[t]
\includegraphics[scale=0.45]{%
\slaninafigdir/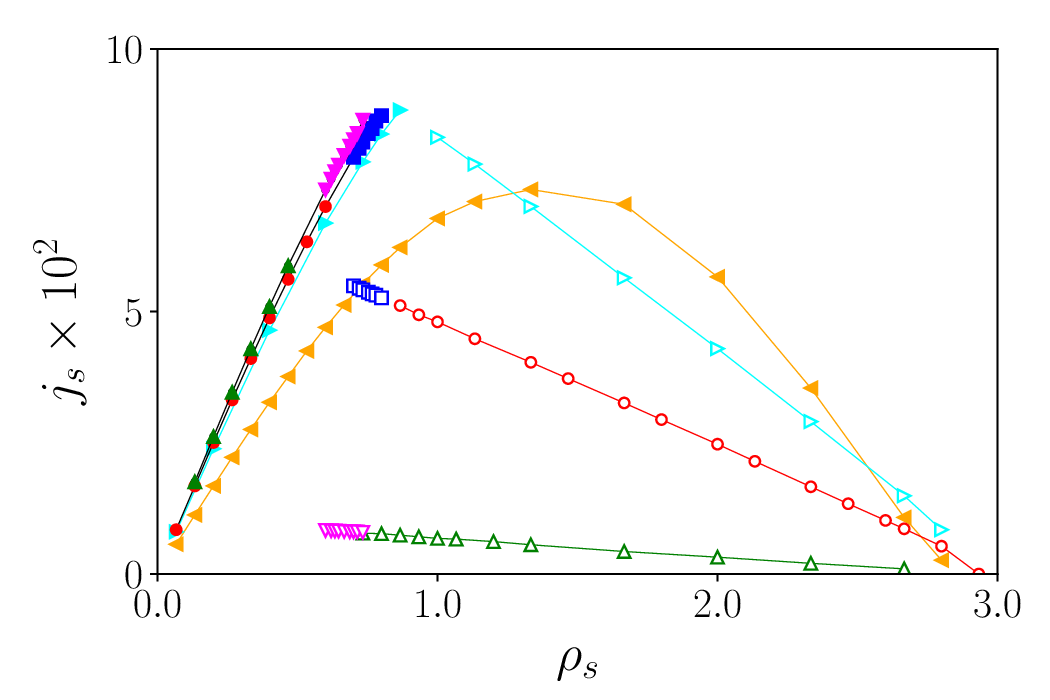}
\caption{Dependence of the current of small particles on the density
of small particles. The length of the system is $L=1000$, the number of big particles is
$N_b=500$.  The switching rate is $\alpha=0.1$ (symbols
{\large$\blacktriangleleft$}),  $\alpha=0.02$ (symbols
{\large$\triangleright$},  {\large$\blacktriangleright$}), $\alpha=0.01$,
(symbols
{\Large$\circ$}, {\Large $\bullet$}, $\square$, $\blacksquare$)
and $\alpha=0.001$, (symbols $\bigtriangleup$, {\large$\blacktriangle$}
$\bigtriangledown$, {\large$\blacktriangledown$}).  The symbols
$\blacksquare$ and {\large$\blacktriangledown$} denote metastable
current before jamming occurs, the symbols  $\square$ and  $\bigtriangledown$
denote
stationary current after jamming.
}
\label{js-vs-rhos-L1000-Nb500-alpha0p01alpha0p001}
\end{figure}

Let us focus first on the
current of small particles only.
We show the
current-density diagram for various values of the parameters in
Figs. \ref{js-vs-rhos-L1000-Nb0Nb2000-alpha0p01} and
\ref{js-vs-rhos-L1000-Nb500-alpha0p01alpha0p001}. Let us look at the
low-density branch of the diagram first. We can observe that it looks
like a segment of current-density
diagram of generalized ASEP model \cite{hum_kot_net_sla_20}. For low
values of the switching rate, about $\alpha \lesssim 0.02$, the shape
of this branch only weakly depends on $\alpha$ and on the number of
big particles $N_b$. Especially, the slope at zero density, which is
the ratchet velocity for free particles, is completely independent of
$N_b$ for reasons explained earlier and remains close to its adiabatic
value as long as  $\alpha \lesssim 0.02$, as was seen in
Fig. \ref{velocity-vs-alpha-L1000-Nb0}.

\begin{figure}[t]
\includegraphics[scale=0.45]{%
\slaninafigdir/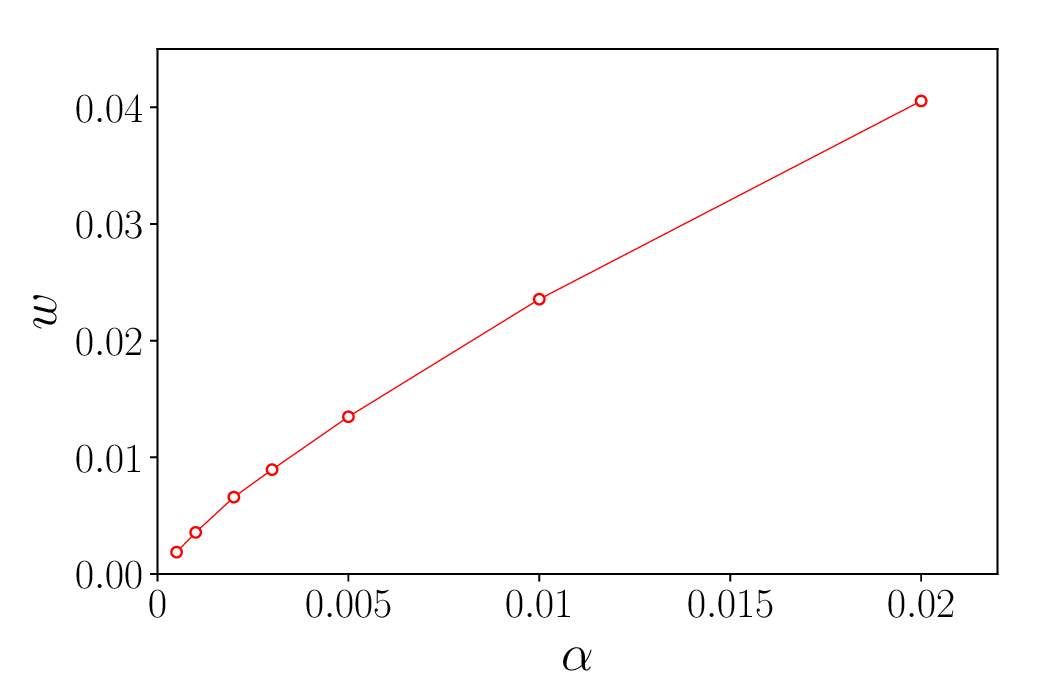}
\caption{Dependence of the slope of the jammed branch of the
current-density diagram on the switching rate.
The length of the system is $L=1000$, the number of big particles is
$N_b=500$.}
\label{slope-jammed-vs-alpha-L1000-Nb500}
\end{figure}

On the
contrary, the high-density branch of the diagram looks very different
from the generalized ASEP. For low switching rate (again, for about
$\alpha \lesssim 0.02$) this branch decreases linearly with density,
$  j_s=u-w \rho_s$,
reaching zero at maximum density allowed by the condition
$\rho_s+2\rho_b\le 3$. The number of big particles $N_b$ determines
the position of the zero, but does not seem to influence the
slope. Therefore, we found that the slope $w$ depends only on the
switching rate, and increases with $\alpha$ as shown in
Fig. \ref{slope-jammed-vs-alpha-L1000-Nb500}.

For switching rates larger than about $0.02$ the current-density
changes its character.
We already mentioned that the notions of
jamming time, as well as those of
metastable and stationary currents start to be ill-defined. Therefore,
we are not able to distinguish the two separate branches of the
diagram and the shape of the current-density diagram approaches the
typical form of generalized ASEP model. In
Fig. \ref{js-vs-rhos-L1000-Nb500-alpha0p01alpha0p001} we show an
example for $\alpha=0.1$. As a result, the high-density part  of the
current-density diagram is no
more linear and therefore also the notion of the slope $w$ looses sense. This is
also the reason why no point beyond $\alpha=0.02$ are given in Fig.
\ref{slope-jammed-vs-alpha-L1000-Nb500}. Overall, the behavior of the
system for switching rates above $\alpha\simeq 0.02$ is just a
somewhat deformed behavior of   generalized ASEP model.

Let us  return to the more interesting region of
low switching rates $\alpha \lesssim0.02$.
The data suggest the following scenario for the current of small
particles in the jammed regime.
The jam, caused by dynamical freezing
of active small particles, is kept compact by particles whose active
drift points toward the bulk of the jam. The current is due to the
particles outside the jam.  We naturally expect the density inside the
jam is close to its maximum value, i. e. $\rho_{s}^{\mathrm{jam}}\simeq
3$.
We also assume that the density outside the jam
$\rho_{s}^{\mathrm{unjam}}$ is only weakly
dependent on the average density $\rho_s$ and the density
$\rho_{s}^{\mathrm{unjam}}$ falls into the regime of free flow, therefore
we can determine the unjammed current
$j_{s}^{\mathrm{unjam}}\equiv j_s(\rho_{s}^{\mathrm{unjam}})$.
On such condition, the number of
particles outside the jam, participating in the current, decreases
linearly with $\rho_s$. Therefore, the current of small particles in
the jammed regime follows
\begin{equation}
j_s=j_{s}^{\mathrm{unjam}}\frac{
\rho_{s}^{\mathrm{jam}}-\rho_s}{
\rho_{s}^{\mathrm{jam}}-\rho_{s}^{\mathrm{unjam}}}\;.
\label{eq:lineardecrease}
\end{equation}

In the data obtained from simulations, the two branches (the free and
the jammed) of the current-density
diagram are disjoint, but we may hypothesize about the point where
they would cross, if the linear regime of the jammed phase would be
linearly extrapolated to lower densities $\rho_s$, outside its actually observed
range. According to (\ref{eq:lineardecrease}), the intersection would
occur at density $\rho_{s}^{\mathrm{unjam}}$, where the value of the
current is $j_{s}^{\mathrm{unjam}}$. This gives more precise meaning to
the quantity we denoted as density outside the jam, which would
otherwise admit rather ambiguous definitions. It also suggests the
interpretation of the density outside the jam $\rho_{s}^{\mathrm{unjam}}$
as the limit of stable stationary free (unjammed) flow. For average
densities $\rho_s > \rho_{s}^{\mathrm{unjam}}$ the free flow is unstable
with respect to formation of the jam. Further addition of small
particles results in increased jam size, but the density outside the
jam remains  $\rho_{s}^{\mathrm{unjam}}$. Such phase coexistence is again
analogous to the situation at first-order phase transition in
equilibrium systems.

This argument also explains why the introduction of big particles
leads just to the leftward shift of the linear dependence of the
current on $\rho_s$, without change of the slope, as shown in
Fig. \ref{js-vs-rhos-L1000-Nb0Nb2000-alpha0p01}. The big particles can
be seen as randomly placed obstacles, whose main effect is diminishing
the volume available for small particles. Therefore, their effect can
be approximately accounted by replacing $\rho_s\to\rho_s+2\rho_b$ in
(\ref{eq:lineardecrease}). The data in
Fig. \ref{js-vs-rhos-L1000-Nb0Nb2000-alpha0p01} support this picture
very well (the slopes for $N_b=0$ and $N_b=2000$ differ by $2\%$ only).


\section{Drag imposed on passive particles}

\begin{figure}[t]
\includegraphics[scale=0.45]{%
\slaninafigdir/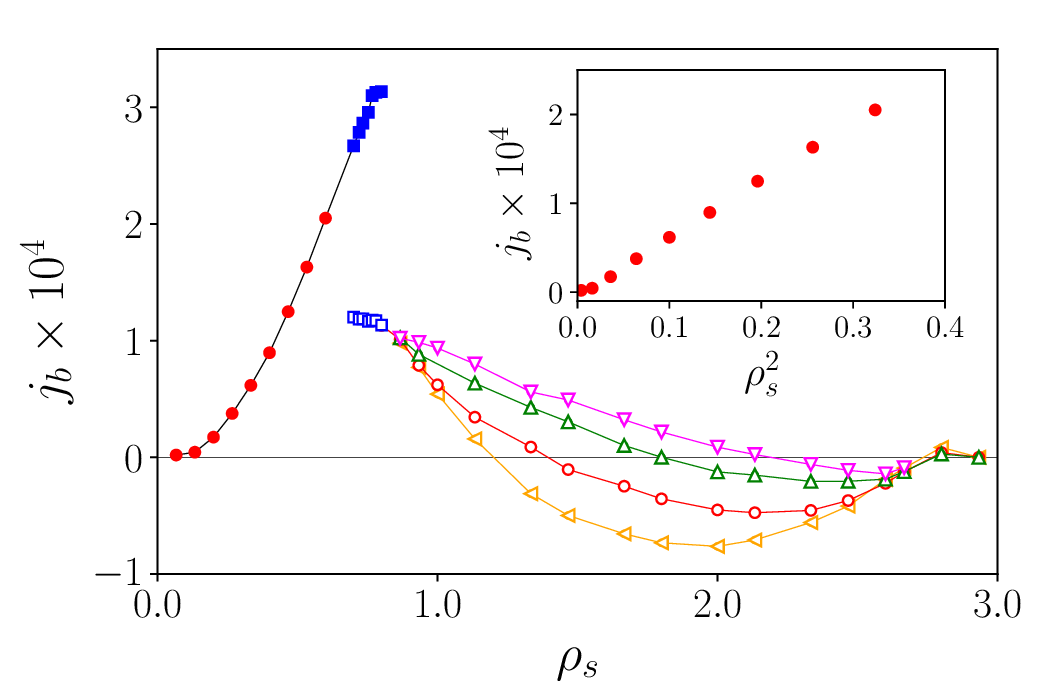}
\caption{Dependence of the current of big particles on the density
of small particles. The length of the system is $L=1000$, the number of big particles is
$N_b=500$, he switching rate is $\alpha=0.01$.
The symbols {\Large $\bullet$} and $\blacksquare$ denote current in
unjammed state. In the jammed state, the average current is taken at
time $t=10^7$ ({\Large$\triangleleft$}), $t=5\cdot 10^7$ ({\Large
$\circ$}), $t=5\cdot 10^8$ ($\bigtriangleup$), and $t=5\cdot 10^9$
($\bigtriangledown$). The symbols $\blacksquare$ denote metastable
current before jamming occurs, the symbols  $\square$ denote
stationary current after jamming. In the inset, detail of the same
data in the region of small densities.
}
\label{jb-vs-rhos-L1000-Nb500-alpha0p01}
\end{figure}

\begin{figure}[t]
\includegraphics[scale=0.45]{%
\slaninafigdir/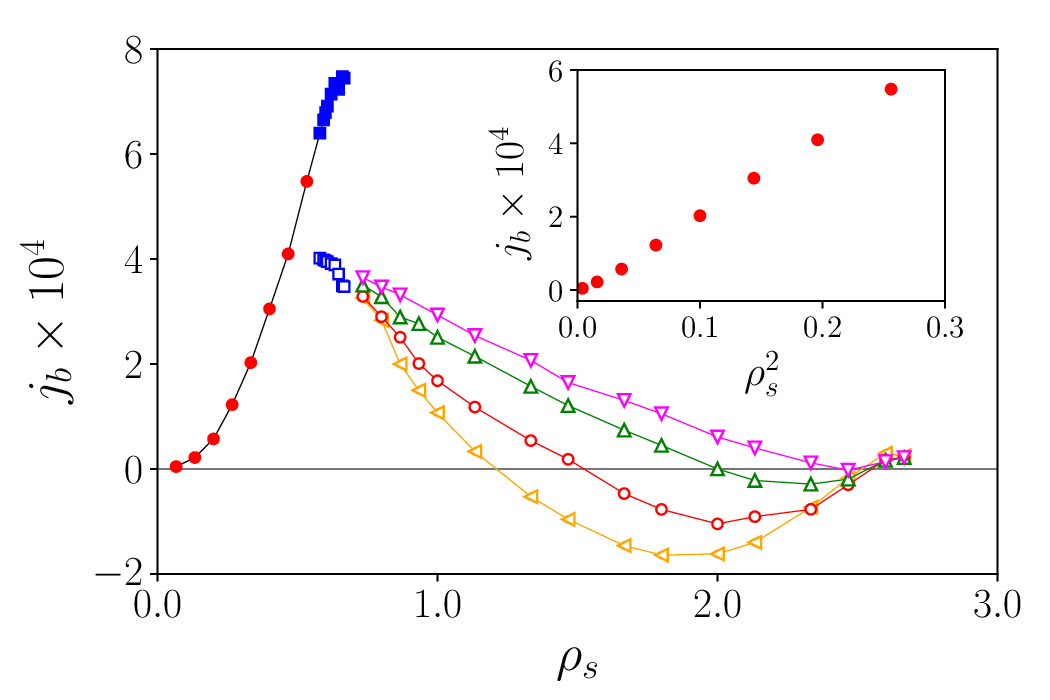}
\caption{Dependence of the current of big particles on the density
of small particles. The length of the system is $L=1000$, the number
of big particles is
$N_b=2000$, he switching rate is $\alpha=0.01$. The meaning of the
symbols is the same as in
Fig. \ref{jb-vs-rhos-L1000-Nb500-alpha0p01}. In the inset, detail of the same
data in the region of small densities.
}
\label{jb-vs-rhos-L1000-Nb2000-alpha0p01}
\end{figure}

Now let us look at  what happens with the passive (big)
particles. From the point of view of active (small) particles, they are just
 {randomly moving}
obstacles that reduce the available space. However, we have already seen in
Fig. \ref{configurations-L1000-Ns8700-Nb2000-alpha0p001}, as well as
in Fig. \ref{evolution-jamming-L1000-Ns8700-Nb2000-alpha0p01},
that the passive particles also  {exhibit rectified movement},
although with velocity which is
several orders of magnitude smaller than the average velocity of active
particles. We can also see that the time
evolution of the velocity
of their movement follows closely the changes of the velocity of
small particles, namely the appearance of a jam is reflected
in the current of both active and passive particles
simultaneously. Also for passive particles we can identify the metastable
and stationary current, as shown in
Fig. \ref{configurations-L1000-Ns8700-Nb2000-alpha0p001}c. As
the passive particles cannot exhibit themselves any ratchet effect,
their current is due to interactions with active particles. A
simplified view can expect that moving active particles exhibit a drag
on the passive particles, much like moving molecules in a flowing
liquid impose a drag on colloid particles suspended in the
liquid. However, it is only partially true and the situation is more
complicated, as shown in Figs. \ref{jb-vs-rhos-L1000-Nb500-alpha0p01}
and \ref{jb-vs-rhos-L1000-Nb2000-alpha0p01}.


The current-density diagram for the current of big particles also
exhibits two branches corresponding to free-flow and jammed regimes as
does the current-density diagram for small particles.
To avoid confusion, let us recall that these branches include also the interval of
densities where both metastable and stationary currents are
measured. In such cases, values of the metastable current belong to
free-flow branch, while the values of stationary flow belong to jammed branch.
In the free-flow branch, the current of big particles grows
quadratically with density of small particles, contrary to the current of small
particles, where the dependence is linear. The quadratic dependence is
demonstrated in the insets in   Figs
\ref{jb-vs-rhos-L1000-Nb500-alpha0p01} and
\ref{jb-vs-rhos-L1000-Nb2000-alpha0p01}.

The behavior in the jammed branch is more complicated from the point
of view of big particles than it was with small particles. After the
formation of the jam, as shown in detail in
Fig. \ref{evolution-jamming-posunute-L1000-Ns10700-Nb500-alpha0p001},
the current of small particles stabilizes at certain value and
fluctuates around it for the rest of the simulation.
For big particles, the same is observed for those densities for which
metastable current is observed. However, for such large densities of
small particles that metastable state is absent, or at least is
shorter than initial transient regime dominated by initial conditions,
the current of big particles behaves in more complicated way. First it
drops to a small value, which may be even negative, i.e. the current
of big particles may go in opposite direction than the current of
small particles. Then, the current of big particles slowly grows and
eventually becomes positive. This suggests that there are two
different regimes, the first regime characterized by the presence of metastable
states and the second, at larger densities, characterized by absence
of metastable states and slow dynamics of jammed states. The first
regime bears the characteristics of thermally activated process,
while the second is typical for spinodal decomposition. We shall
return to interpretation of these regimes later.

We illustrate the slow increase of the current of big particles in
Figs  \ref{jb-vs-rhos-L1000-Nb500-alpha0p01} and
\ref{jb-vs-rhos-L1000-Nb2000-alpha0p01} by showing the value of the
current of big particles at different times, namely at times
$t=10^7,\;5\cdot 10^7,\; 5\cdot 10^8,\; 5\cdot 10^9$. We can see
monotonous increase with time for densities in the interval
$\rho_s\in[\rho_{s-},\rho_{s+}]$. The lower edge of the interval
apparently coincides with the maximum density at which we are able to
see the metastable current, while the upper edge is only slightly
below the maximum density allowed by the exclusion principle.
The
increase of the current $j_b$ is most pronounced for
densities in the middle of this range. As we already mentioned, during the
evolution the current of big particles may even change sign, for
example for density $\rho_s=1.5$ and densities close to it. This
implies that just after the jam is formed, the current of small
particles imposes a drift
of opposite direction on the big particles, but this drift
gradually diminishes in absolute value, then changes sign and keeps
growing in the same direction as the current of small particles.

\begin{figure}[t]
\includegraphics[scale=0.45]{%
\slaninafigdir/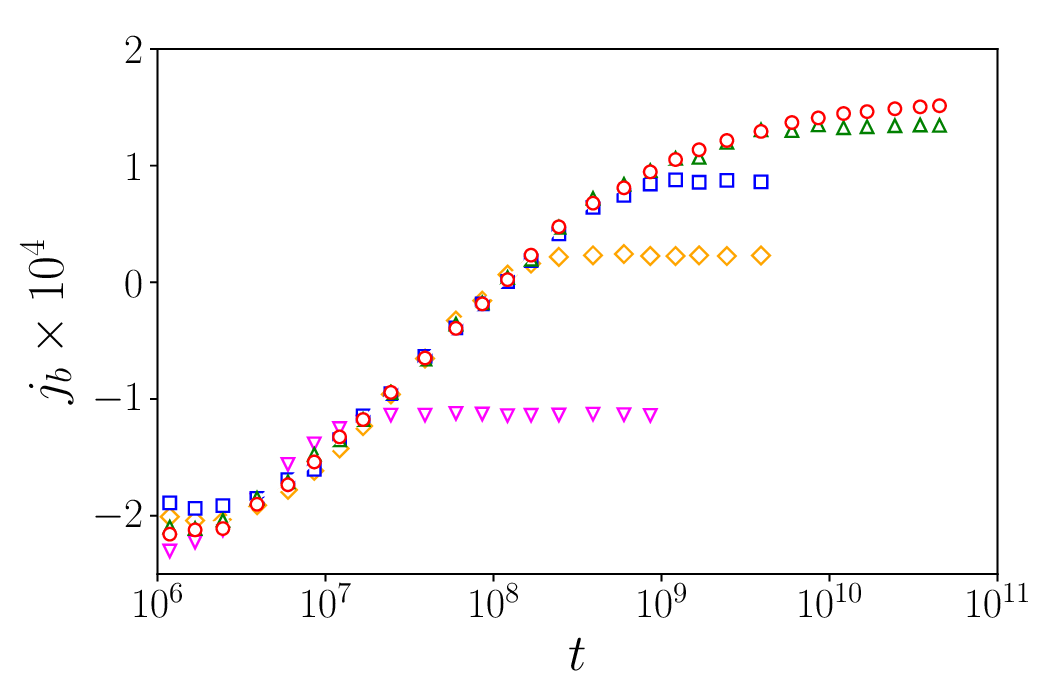}
\caption{Time evolution of the average current of big particles in a
jammed state. The system size is $L=1000$ ({\Large$\circ$}), $L=500$
($\bigtriangleup$), $L=200$ ($\square$), $L=100$
({\Large$\diamond$}), and $L=50$ ($\bigtriangledown$). The number
of small  particles is $N_s=25\,L$, the number of big
particles   is $N_b=2\,L$. The switching rate is
$\alpha=0.01$.
}
\label{jb-evol-rho25}
\end{figure}

We show in Fig. \ref{jb-evol-rho25} time dependence of the current of
big particles for density of small particles $\rho_s=5/3$ and density
of big particles $\rho_b=2/15$. In order to assess the finite-size
effects, we plot the results for sizes
$L=50,\; 100,\; 200,\; 500,\; 1000$.
We can clearly see the change of direction, in this case around time
$t\simeq 10^8$. The increase is extremely slow, at beginning
logarithmic in time, later it seems even slower. At the largest system
sizes accessible in this simulation ($L=1000$) we are still not sure
whether the current reaches truly stationary value, but for smaller
sizes ($L=500$ and smaller) we observe that after some time the
current settles at a saturated value. The saturated value is positive
for $L=100,\;200,\; 500$, but is negative for $L=50$. This suggests
that for small enough system the inversion of current direction never
takes place. The data we were able to collect do not allow to make
firm statements about the behavior in thermodynamic limit
$L\to\infty$. However, the data suggest that for large enough system,
the true stationary current of big particles is always positive,
although the quasi-stationary current just after formation of the jam
and for certain time afterward may be negative. In any case, the slow
evolution of the current shown in Fig.  \ref{jb-evol-rho25} is a sign
of very slow (logarithmic) relaxation and reconstruction of the
stationary configuration of the mixture of small and big particles.

\begin{figure}[t]
\includegraphics[scale=0.45]{%
\slaninafigdir/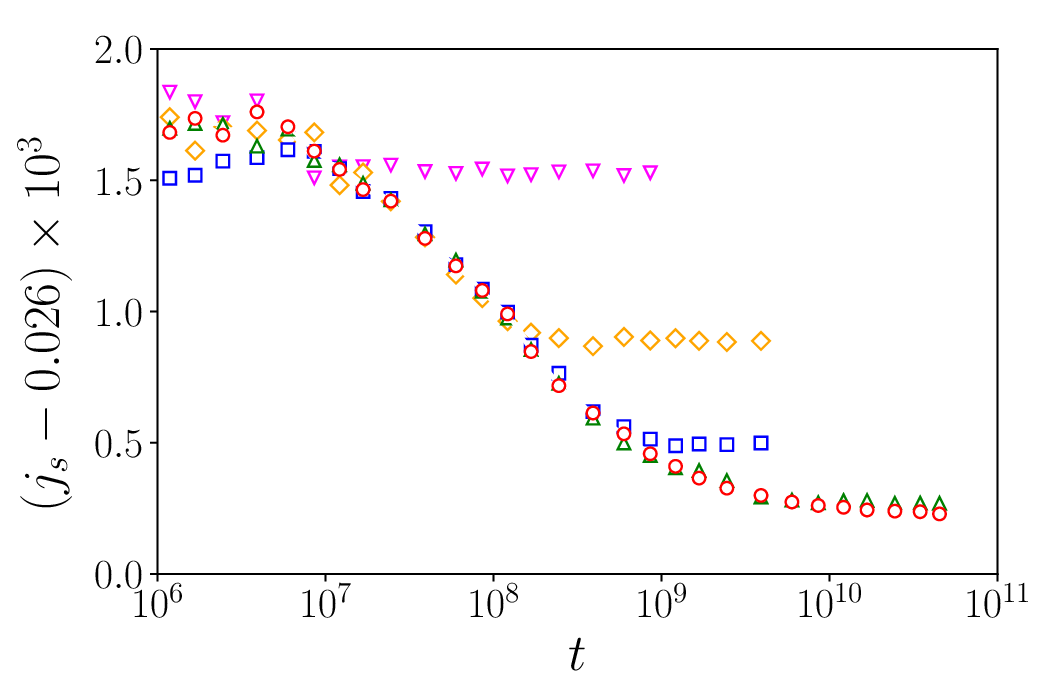}
\caption{Time evolution of the average current of small particles in a
jammed state. The system size is $L=1000$ ({\Large$\circ$}), $L=500$
($\bigtriangleup$), $L=200$ ($\square$), $L=100$
({\Large$\diamond$}), and $L=50$ ($\bigtriangledown$). The number
of small particles is $N_s=25\,L$, the number of big
particles   is $N_b=2\,L$. The switching rate is
$\alpha=0.01$.
}
\label{js-evol-rho25}
\end{figure}

It is also interesting to look at backward influence of such slow
relaxation on the behavior of small particles. In parallel with the
time evolution of the current of big particles, we show in Fig.
\ref{js-evol-rho25} the time evolution of the current of small
particles. We can see that the small particles slow down while the
current of big particles increases and both currents saturate at the
same time at a true
stationary value (actually observed for $L\le 500$ and expected for
larger sizes). However, the decrease of
$j_s$ is relatively much smaller than the change in $j_b$. In Fig.
\ref{js-evol-rho25} we can see that during the slow relaxation the
value of $j_s$ drops by less than $10\%$.

\begin{figure}[t]
\includegraphics[scale=0.45]{%
\slaninafigdir/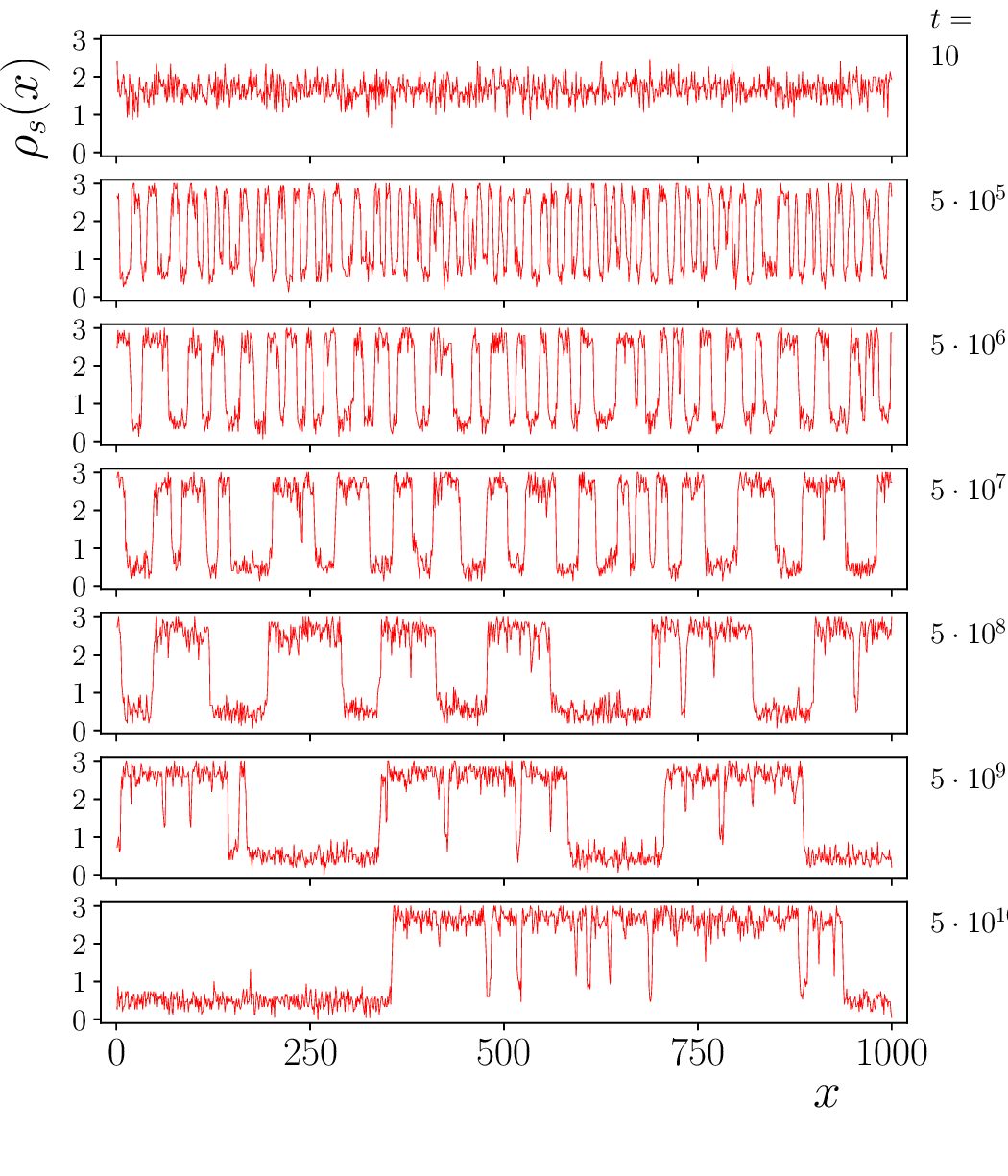}
\caption{Snapshots of the local density of small particles, at times
indicated on the right side of the panels.
The system size is $L=1000$, the number
of small particles is $N_s=25000$, the number of big
particles   is $N_b=2000$. The switching rate is
$\alpha=0.01$. The local density $\rho_s(x)$ is the average density
of small particles in the $x$-th tooth.
}
\label{clusters-evol-rho25}
\end{figure}

In order to better understand the slow relaxation of the current,
we looked at the
evolution of the configuration of particles. We show in
Fig. \ref{clusters-evol-rho25} snapshots of the local density of small
particles, for the same density and switching rate as used in
Figs. \ref{jb-evol-rho25} and \ref{js-evol-rho25}. In the sequence of
snapshots we observe that jamming starts by forming large number of
small jamming clusters which gradually grow and coalesce. The true
stationary state is reached only after
all of the clusters aggregate into a single cluster, which then keeps slowly
moving as a whole, but does not increase in size. This is a
very slow process and this determines the slow approach of the current
to the stationary value. We looked also at snapshots of the local density of
big particles (not shown), but this quantity does not manifest traces
of the jamming clusters. Any possible correlation between the density
of small and big particles is, at least in the snapshots,
overshadowed by stochastic fluctuations. However, even the qualitative
picture we can observe from Fig. \ref{clusters-evol-rho25} illustrates
the interpretation of the slow dynamics as evolution governed by
spinodal decomposition, which is in force for densities within the
interval $\rho_s\in[\rho_{s-},\rho_{s+}]$. The clusters grow by
coalescence. On the contrary, for lower densities, $\rho_s<\rho_{s-}$
the clusters appear and grow by nucleation event, as illustrated in
Fig. \ref{configurations-L1000-Ns8700-Nb2000-alpha0p001}.

To summarize, we can observe several distinct regimes in the
current-density diagram, as function of the density of small
particles. For small density, $\rho_s<\rho_{sm}$ no jamming occurs and
the current depends on density as in a variant of generalized ASEP
model. At density $\rho_{sm}$, which is however difficult to establish
precisely in the simulations, the metastable states first occur and
the current-density diagram is characterized by two distinct values of
current, the larger being the metastable current, the lower the
current after formation of the jam. Such double-valuedness persists in
the interval $\rho_{sm}<\rho_s<\rho_{s-}$. In this interval, the
dynamics is essentially thermally activated. At the density
$\rho_{s-}$ uniform state becomes unstable and decays by spinodal
decomposition. The dynamics is logarithmically slow and proceeds by
coalescence of jammed clusters. This holds in the interval
$\rho_{s-}<\rho_s<\rho_{s+}$, where the upper spinodal
$\rho_{s+}$ is so close to the maximum density that for larger
densities the systems looks
jammed already from the beginning of the evolution.
In fact, this behavior corresponds to the scenario of spinodal
decomposition described by phenomenological equations
\cite{cat_tai_15} as well as in exact hydrodynamics of a simple
active exclusion process \cite{kou_eri_bod_tai_18}.


\section{Conclusions}

We investigated by numerical simulations a lattice model of a mixture
of passive and active particles in a pore whose aperture varies
periodically along its axis. The active drift of
the active particles can point in one of four directions, which is
changed with switching rate $\alpha$.
The steric repulsion of the particles is
accounted for by an exclusion constraint, limiting the number of
particles of each type in each cell of the lattice. Due to the internal
driving of the active particles, a ratchet effect occurs. This implies
non-zero average current of the active particles, which in turn
imposes a drag on passive particles, whose average current is therefore
also non-zero.

In the limit of
vanishing density of active particles the ratchet velocity is the
function of the switching rate only. In our model, the dependence of
this velocity on the density of passive particles is absent by
construction, although in more general models such dependence can
appear. In the limit $\alpha\to 0$ this ratchet velocity approaches an
adiabatic value, while for $\alpha\to\infty$ it decreases as
$1/\alpha$. This behavior is consistent with earlier studies of
non-interacting active particles.

When the density of active particles increases, new effects appear,
especially the dynamical freezing, or motility induced phase
separation. In our pore geometry, it manifests
itself as dynamical jamming transition, where the stationary ratchet
current drops to appreciably lower value, although not to zero. At
sufficiently low switching rate, $\alpha\lesssim 0.02$, the jamming
follows rather simple scenario. The system remains for certain time in
a quasi-stationary or metastable state with no dynamical
freezing. Then, at a
certain moment which we call jamming time, a frozen cluster emerges,
with local density of particles nearly the maximum value permitted by
the exclusion principle. This cluster steadily grows, until it reaches
size determined by average particle density. On the other hand, the
density outside the
cluster depends only on the switching rate and is independent of the
average density.
The jamming times have exponential probability
distribution and the average jamming time is itself an exponential
function of the density of active particles.

Practically, the jamming is clearly observable only in certain
interval of densities of active particles. Within such interval, two
characteristic values of current are attributed to a single value of
density, namely the metastable current before the jam appears
and the asymptotic current after formation of the jam. The
current-density digram has therefore two separate branches.
For smaller densities the
jam never appears during accessible simulation time and the observed
current is continuation of the branch of metastable currents.
On the other side for larger
densities the jam appears so early that the jamming time is comparable
or even shorter than the initial transient time during which the
system remembers the starting configuration (which is completely
random placement of particles). Therefore, we observe only the
stationary current in jammed configuration and in the current-density
diagram we draw the continuation  of the branch of stationary current.
The form of the current-density diagram characterized by two separate
branches partially overlapping in the middle but never crossing, is
observed for $\alpha\lesssim 0.02$ for all densities of passive
particles and for both the current  of active and current of passive
particles.

The low-density branch has the form which closely resembles a segment
of the current-density diagram characteristic of generalized ASEP
models. It depends very weakly on the switching rate and the density
of passive particles. Interestingly, for
small densities of active particles the current of
passive particles increases quadratically with the density of active
particles, while  the current of active
particles increases linearly as usual in ASEP-like models.

The high-density branch behaves in qualitatively different
manner. Most importantly, the current of active
particles decreases linearly with the density of active particles in
the whole interval of
densities for which the branch is defined and approaches zero at the
maximum density allowed by the exclusion constraint. The slope of this linear
dependence increases (in absolute value) with the switching
rate. This finding implies the following scenario of the jamming
transition. The local density of active particles within the jam is
always close to
its maximum allowed by the exclusion constraint, which is $3$ in our
model. The density  of active particles outside the jam has lower
value, which depends only on the switching rate but not on the average
density of active particles.
This picture is reminiscent of phase coexistence in equilibrium
systems with first-order phase transition. Also the very existence of
long-lived metastable unjammed states is consistent with such
view. This regime corresponds to thermally activated process, where
clusters are formed by nucleation.

The behavior of the high density branch in the current of passive
particles is even more complicated. After the jam is formed, we
observed very slow approach of
the current of passive particles to a
stationary state, logarithmic in time.
The current of passive particles can
even become negative, i.e. the passive particles are pushed in the
opposite direction than the active particles flow. However, during the
slow evolution the direction changes and the ultimate current of
passive particles has the same direction as the current of active
particles. The data also suggest that the true stationary current of
passive particles may follow the same linear dependence on the density of
active particles as does the current of active particles. However,
more data would be necessary to fully prove this hypothesis.

The detailed observation of the particle configurations during this slow
evolution shows that the clue consists in slow dynamics of jamming
clusters 
which
corresponds to spinodal decomposition of an
unstable homogeneous state. This is in contrast to the nucleation process
which is at work for lower densities and which leads on one side to
long-lived metastable states, but on the other hand avoids the very
slow dynamics after the jamming event.

Overall, the observed current-density diagrams can distinguish four
regimes, depending on the density of small particles. In the
low-density regime the unjammed homogeneous state is stable. At higher
densities, the unjammed state is metastable and has finite
lifetime. Two characteristic values of the current are attributed to a
single
density, namely the metastable current and the stationary one. This behavior
persists until the density reaches the lower spinodal, where
homogeneous state becomes unstable. For densities from the lower
spinodal up to the upper
spinodal the dynamics is logarithmically slow. The upper spinodal is very close
to maximum possible density and beyond it the system behaves as jammed
from the beginning. This behavior corresponds well to spinodal
decomposition observed in exact hydrodynamic limit of a simple
one-dimensional model
of active lattice gas \cite{kou_eri_bod_tai_18}.

In this work, we concentrated mainly on the low-switching-rate regime,
$\alpha\lesssim 0.02$, which we consider more physically interesting.
In fact, for larger switching rates, $\alpha\gtrsim 0.02$, the
behavior of the system
becomes
much less clear-cut.
Recall that our interpretation of the simulation results relied mainly
on the view of jammed configurations as infinitely deep traps.
We found that for $\alpha\gtrsim 0.02$ the traps become effectively
shallow, enabling dissolution of the jams. Therefore, the jams have
finite lifetime and the current averaged over long enough time can be
viewed as weighted average of the currents in the jammed and unjammed
state. Overall, the current-density diagram closely resembles that of
the generalized ASEP models and looses the specific two-branch
character, which is distinctive for low switching rates.

Clearly, the threshold value $\alpha\simeq 0.02$ separating the
regimes is to large extent dictated by the computer time we devoted to
our simulations. Most probably, for much longer runs we would observe
escape from the traps, i.e. dissolution of the jams, even for lower
$\alpha$. However, we can also look at the problem from the opposite
side. Accepting that the jams have principally finite lifetime, we can
include this lifetime in the set of quantities characterizing the
system, for however low $\alpha$ we have. Moreover, even with finite
lifetime of the jams, we can in
principle measure the value of the current only when in jammed
state and also the current only at times when there is no jam. This
way we could get a two-valued current-density dependence also for
larger $\alpha$. Therefore, the value $\alpha\simeq 0.02$ we used as a
threshold separating two regimes is rather conventional and set by convenience.

On the other hand, it is also possible that there is a true critical
value $\alpha_c$ above which the two-branch structure of the
current-density diagram disappears for physical reasons. If it exists,
it would be certainly larger than our conventionally set
$\alpha\simeq 0.02$. Let us
formulate at least one argument in support of such hypothesis.
The typical form of the current-density diagram in generalized ASEP
models is a concave function with a single maximum and linear approach
to zero at both ends, i. e. for densities close to zero and close to
maximum given by exclusion constraint. First, we can reasonably assume that
this same form would hold also in our model with active particles on
condition that the formation of the jam is somehow
inhibited. Therefore, that would be the form of the density dependence
of the metastable current, if it was observable in the whole range of
densities. At the maximum density, the curve approaches zero at slope
$w_\mathrm{max}$. Second, we also assume, although now with lower
level of certainty, that we can generalize to all densities and all
switching rates the observation we made for low $\alpha$, namely that
the local density inside the jam is always nearly
the maximum one, while outside the jam the local density depends only
on switching rate and not on the average density. The latter
assumption leads to linear decrease in the branch of jammed
current. However, this branch lies always lower than the curve of
metastable currents.  Therefore,the slope $w$ of the jammed branch cannot
exceed the maximum $w_\mathrm{max}$. We have seen that the slope $w$
increases with $\alpha$ and moreover, the increase is approximately
linear for large enough $\alpha$. Therefore, it is reasonable to
assume that at certain $\alpha_c$ the slope hits its maximum
$w_\mathrm{max}$. This would fix the critical point $\alpha_c$ of the
jamming transition in our system. Of course, we cannot exclude another
scenario, that the increase of $w$ with $\alpha$ slows down and does
not reach $w_\mathrm{max}$ for any finite $\alpha$. In this case the
critical point would be absent. However, we consider this second
scenario less plausible.

Supposing that the finite critical point  $\alpha_c$ does exist, the
distinction between the regimes below and above the threshold
$\alpha\simeq 0.02$  gains more deep meaning beyond the pragmatic view
we kept in the interpretation of our simulation data. Physically, the
behavior we observed for $\alpha\lesssim 0.02$ is the typical picture of
what happens  in the limit of infinite sizes and times in the
sub-critical phase   $\alpha <\alpha_c$, while the behavior at
$\alpha\gtrsim 0.02$  is the picture of the super-critical phase
$\alpha>\alpha_c$. The pragmatically obtained threshold
$\alpha\simeq 0.02$ would then serve as a lower estimate of the true
value of the critical point $\alpha_c$. Investigation of the
properties of this critical point, including possible critical
exponents, is a question for future research.

\begin{acknowledgments}
We wish to thank K. Neto{\v{c}}n\'y for inspiring discussions.
Computational resources were provided by the e-INFRA CZ project
(ID:90140), supported by the Ministry of Education, Youth and Sports
of the Czech Republic.
\end{acknowledgments}
%
%
%
%
%
%

%
%
%
%
\end{document}